\documentstyle[subeqnarray,preprint,pre,aps,eqsecnum,epsf,graphicx]{revtex}

\newcommand{\real}{{\sf I}\kern-.12em{\sf R}}
\newcommand{\comp}{{\sf I}\kern-.50em{\sf C}}
\newcommand{\unity}{{\sf I}\kern-.54em{\sf 1}}
\def\spose#1{\hbox to 0pt{#1\hss}}
\def\ltapprox{\mathrel{\spose{\lower 3pt\hbox{$\mathchar"218$}}
 \raise 2.0pt\hbox{$\mathchar"13C$}}}

\begin{document}

\rightline{IFUP-TH 37/04}

\centerline{\bf Analyticity in $\theta$ on the lattice and}
\centerline{\bf the large volume limit of the topological susceptibility}
\vskip 5mm
\centerline{B. All\'es$^a$, M. D'Elia$^{b}$ and A. Di Giacomo$^c$}
\centerline{\it $^a$INFN, Sezione di Pisa, Pisa, Italy}
\vskip 2mm
\centerline{\it $^b$Dipartimento di Fisica, Universit\`a di Genova and INFN,
Genova, Italy}
\vskip 2mm
\centerline{\it $^c$Dipartimento di Fisica, Universit\`a di Pisa and INFN,
Pisa, Italy}

\begin{abstract}
Non--analyticity of QCD with a $\theta$ term at $\theta=0$ may signal
a spontaneous breaking of both parity and time reversal invariance.
We address this issue by investigating the large volume limit of the
topological susceptibility $\chi$ in pure $SU(3)$ gauge theory. 
We obtain an upper bound for the symmetry breaking order parameter 
$\langle Q \rangle$ and, as a byproduct, the value
$\chi=\left(173.4(\pm 0.5)(\pm 1.2)(^{+1.1}_{-0.2})\;{\rm MeV}\right)^4$ 
at $\beta=6$ ($a\approx 0.1$ fermi).
The errors are the statistical error from our data, the
one derived from the value used for $\Lambda_L$ and an estimate of
the systematic error respectively.
\end{abstract}

\vskip 5mm

\vfill\eject

\section{Introduction}

\vskip 5mm

The QCD Lagrangian ${\cal L}_{\rm QCD}$ is parity invariant.
On the other hand the possible spontaneous breaking of parity
is experimentally ruled out to a high precision. There is a debate
as to whether the absence of spontaneous breaking can be analytically
proved in QCD.
By studying the Euclidean formulation of the theory, Vafa and
Witten~\cite{vafa} argued that such breaking is not possible.
Their argument is the following: spontaneous breaking of 
parity occurs when there exists some
parity violating operator ${\cal O}$ whose vacuum expectation value
remains nonvanishing after sending $\theta\rightarrow 0$ in the
extended theory ${\cal L}_{\rm QCD} + \theta {\cal O}$. However
in the Euclidean formulation of the theory, any parity--odd
operator must pick up an imaginary $i$ factor(\footnote{This is not true
in general at finite temperature~\cite{cohen} and the 
possibility of the existence of a stable parity breaking phase at nonzero 
temperature is still open~\cite{kharzeev}.}). Therefore the
$\theta {\cal O}$ term contributes a phase to the Euclidean partition
function and hence the free energy has its minimum at $\theta=0$ and
$\langle{\cal O}\rangle$ vanishes.

As pointed out by several authors~\cite{azcoiti,ji,asorey} the above
argument is insufficient to exclude the scenario of spontaneous
parity breaking. It is necessary to prove as well that the
free energy and its derivatives at its minimum are continuous
functions of $\theta$.

If the parity violating probe is the topological charge density $Q(x)$,
then it is possible to show~\cite{asorey} that the partition
function $Z(\theta)$ is finite throughout the whole complex plane of
$\theta$ and arguments can be given against the appearance of dangerous
Lee--Yang zeros~\cite{leeyang}.

We want to study on the lattice the question about the 
continuity of the derivatives of the free energy for the
case of the parity--odd (and time inversion--odd)
operator $Q(x)$ and give an upper bound to the order
parameter $\langle Q\rangle$ where $Q\equiv\int{\rm d}^4x\, Q(x)$ is
the total topological charge.

We assume that the free energy $E(\theta)$ of
the pure gauge theory at its minimum 
(at $\theta=0$, after Vafa and Witten theorem) has a
discontinuous derivative. If we call
d$E(\theta)/$d$\theta|_{\theta\rightarrow\pm 0}=\pm\alpha$ then
the vacuum expectation value $\langle Q\rangle=\pm V\alpha$ where
$V$ is the spacetime volume and $\alpha$ is a positive real number
which specifies the density of spontaneously generated net topological
charge. The alternating sign indicates the choice of the vacuum.
By inserting a complete set of intermediate states
$\;\unity = |0\rangle\langle 0| + |g\rangle\langle g|+\cdots$
($|g\rangle$ represents pure gluonic states)
between the two topological charge operators in $\langle Q^2\rangle$
we obtain
\begin{equation}
\frac{\left\langle Q^2\right\rangle}{V} = \alpha^2 V + \chi \; ,
\label{defchi}
\end{equation}
where $\chi$ is the usual topological susceptibility
defined in~\cite{witten}. Notice
that if $\alpha\not= 0$ then the above expectation value depends
linearly on the volume.
In the present paper we investigate this volume dependence 
on the lattice for the
pure $SU(3)$ Yang--Mills theory in order to set
a bound on the value of the slope $\alpha$.

\section{The calculation of $\chi$}

\vskip 5mm

The quenched theory was simulated by using the 
standard plaquette action~\cite{wilson}
on the lattice at the inverse bare coupling $\beta=6.0$ on three
volumes: $16^4$, $32^4$ and $48^4$. In all cases a heat--bath
algorithm combined with overrelaxation was used for updating configurations
and care was taken (by checking the autocorrelation function) 
to decorrelate successive measurements in order to
render them independent.

The total topological charge was measured with the
once--smeared operator~\cite{christou} 
$Q_L^{(1)}=\sum_x Q_L^{(1)}(x)$. 
The lattice equivalent of Eq.(\ref{defchi}) is 
$\chi_L\equiv\left\langle \left(Q_L^{(1)}\right)^2\right\rangle /L^4$.
$L^4$ is the dimensionless volume of the lattice, $L^4=V/a^4$, where $a$
is the lattice spacing.

In Figure~1 and Table~1 we show the 
result for $\chi_L$ for the three volumes.
Our investigation requires a precise determination of the topological
susceptibility, hence we used huge statistics. The simulations
were performed with the APEmille facility in Pisa.

In general the lattice topological susceptibility $\chi_L$ is related to
the physical one $\chi$ by a multiplicative and an additive
renormalization~\cite{campostrini}. The equation that expresses this
relationship is

\newpage

\centerline{{\bf Table 1}: Value of $\chi_L$ and statistics for each
lattice size.}
\vskip 1mm
{\centerline{
\vbox{\offinterlineskip
\halign{\strut
\vrule \hfil\quad $#$ \hfil \quad &
\vrule \hfil\quad $#$ \hfil \quad &
\vrule \hfil\quad $#$ \hfil \quad \vrule \cr
\noalign{\hrule}
{L^4} &
{\rm statistics} &
{10^5\;\chi_L} \cr
\noalign{\hrule}
16^4 &
120000 &
1.550(7) \cr
\noalign{\hrule}
32^4 &
60000 &
1.590(9) \cr
\noalign{\hrule}
48^4 &
50000 &
1.580(11) \cr
\noalign{\hrule}
}}
}}

\vskip 2cm

\noindent 
\begin{equation}
\chi_L=\frac{\left\langle\left(Q_L^{(1)}\right)^2\right\rangle}{L^4} =
Z^2 a^8 \alpha^2 L^4 + Z^2 a^4 \chi + M\equiv Z^2 a^4\chi_P + M\; ;
\label{chilat}
\end{equation}
{}for convenience we call 
$a^4\chi_P\equiv a^8 \alpha^2 L^4 + a^4 \chi$.

We renormalize the topological charge operator $Q_L^{(1)}$ by imposing
that it takes integer eigenvalues in the continuum limit. To obtain 
this result we must introduce a renormalization 
constant~\cite{haris,allesvicari} which is finite in virtue of 
the renormalization group invariance of $Q$ in the quenched theory.
This is the origin of the factor $Z$ in expression~(\ref{chilat}).

The operator expansion of the product
$\left(Q_L^{(1)}(x)\,Q_L^{(1)}(0)\right)$ contains a contact
term~\cite{zimmermann}. Part of this term must be 
subtracted
and this is $M$ in expression~(\ref{chilat}). We fix this additive
subtraction by imposing that the topological susceptibility must
vanish in the absence of instantons~\cite{npb}. By construction $M$ is
independent of the background topological sector.

These definitions are valid when $\alpha=0$ and apply
equally well to the case where $\alpha\not= 0$.

To extract information about the parameter $\alpha$
in Eq.(\ref{chilat}) we have to know $Z$ and $M$. 
We have calculated these renormalization
constants paying attention to their possible volume dependence.

\newpage

\begin{figure}
\centerline{\includegraphics[width=\columnwidth]{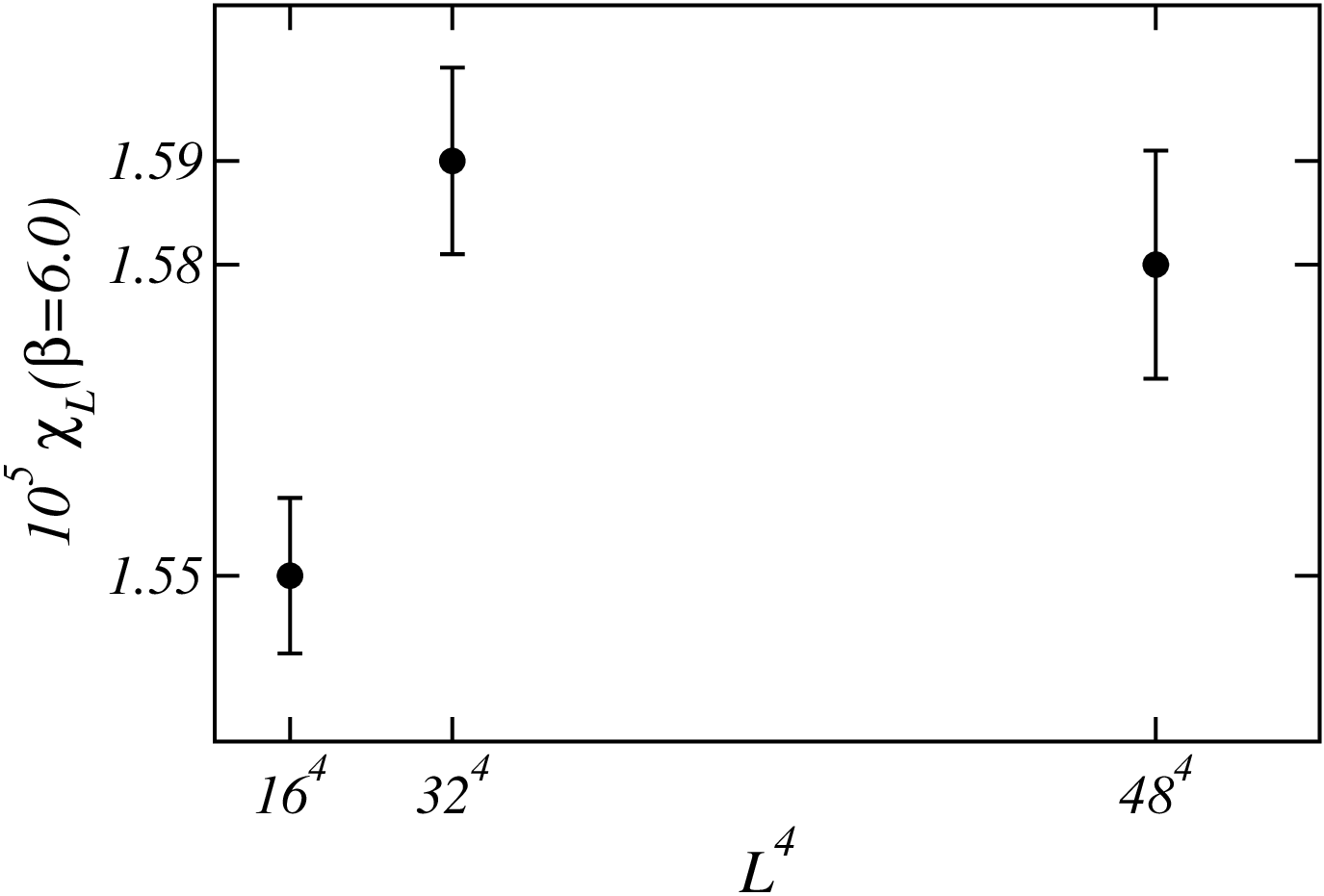}}
\vspace*{12pt}
\caption{$\chi_L$ versus the lattice size for $\beta=6.0$.}
\end{figure}

\section{Calculation of $Z$ and $M$}

\vskip 5mm

We determined the renormalization constants $Z$ and $M$ by 
the nonperturbative method introduced in~\cite{vicaridg,gunduc}.

{}Following the meaning of $Z$, we calculate it by computing
the average topological charge within a fixed topological sector.
If we choose a topological sector of charge $n$ (any nonzero integer) then
\begin{equation}
Z=\frac{\langle Q_L^{(1)}\rangle |_{Q=n}}{n} \; ,
\end{equation}
where the division by $n$ entails the requirement that $Q$ takes
integer values as described above. The brackets
$\langle\cdot\rangle |_{Q=n}$
mean thermalization within the topological sector of charge $n$.

We start our algorithm with a classical configuration with
topological charge 1 ($n=1$) and action $8\pi^2$ in appropriate units.
Then we apply 80 heat--bath updating
steps and measure $Q_L^{(1)}$ every 4 steps. 
This set of 20 measurements is called ``trajectory''.
After each measurement
we cool the configuration to verify that the topological sector
is not changed. We repeat
this procedure to obtain a number of trajectories. For each trajectory we
always discard the first measurements because the configuration is not
yet thermalized. Averaging over the thermalized steps
(as long as the corresponding cooled configuration shows the correct
background topological charge, $n=1$ within a deviation $\delta$)
yields $\langle Q_L^{(1)}\rangle |_{Q=1}$. We estimate
the systematic error that stems from the choice 
of $\delta$ as in~\cite{npb}.

We followed the above procedure on three lattice sizes to study any
possible volume dependence of $Z$. In Table~2 the number of trajectories
and volume sizes are displayed.

\vskip 1cm

\centerline{{\bf Table 2}: Number of trajectories utilized in the
calculation of $Z$ and $M$.}
\vskip 1mm
{\centerline{
\vbox{\offinterlineskip
\halign{\strut
\vrule \hfil\quad $#$ \hfil \quad &
\vrule \hfil\quad $#$ \hfil \quad &
\vrule \hfil\quad $#$ \hfil \quad \vrule \cr
\noalign{\hrule}
{L^4} &
{Z} &
{M} \cr
\noalign{\hrule}
8^4 &
147000 &
129000 \cr
\noalign{\hrule}
12^4 &
60000 &
74000 \cr
\noalign{\hrule}
16^4 &
50000 &
57000 \cr
\noalign{\hrule}
}}
}}

\vskip 2cm

In Figure~2 we show the results for $Z$ extrapolated to
the form $A + B/L^2$. They look very stable, mainly at
the lattice sizes of our interest ($L=16$, $32$ and $48$).
The errors were calculated including the cross
correlation, $\langle A B\rangle - \langle A\rangle \langle B\rangle
\approx -6\times 10^{-5}$.
The $\chi^2/{\rm d.o.f.}$ test leads to~0.02.

\vskip 2cm

\begin{figure}
\centerline{\includegraphics[width=\columnwidth]{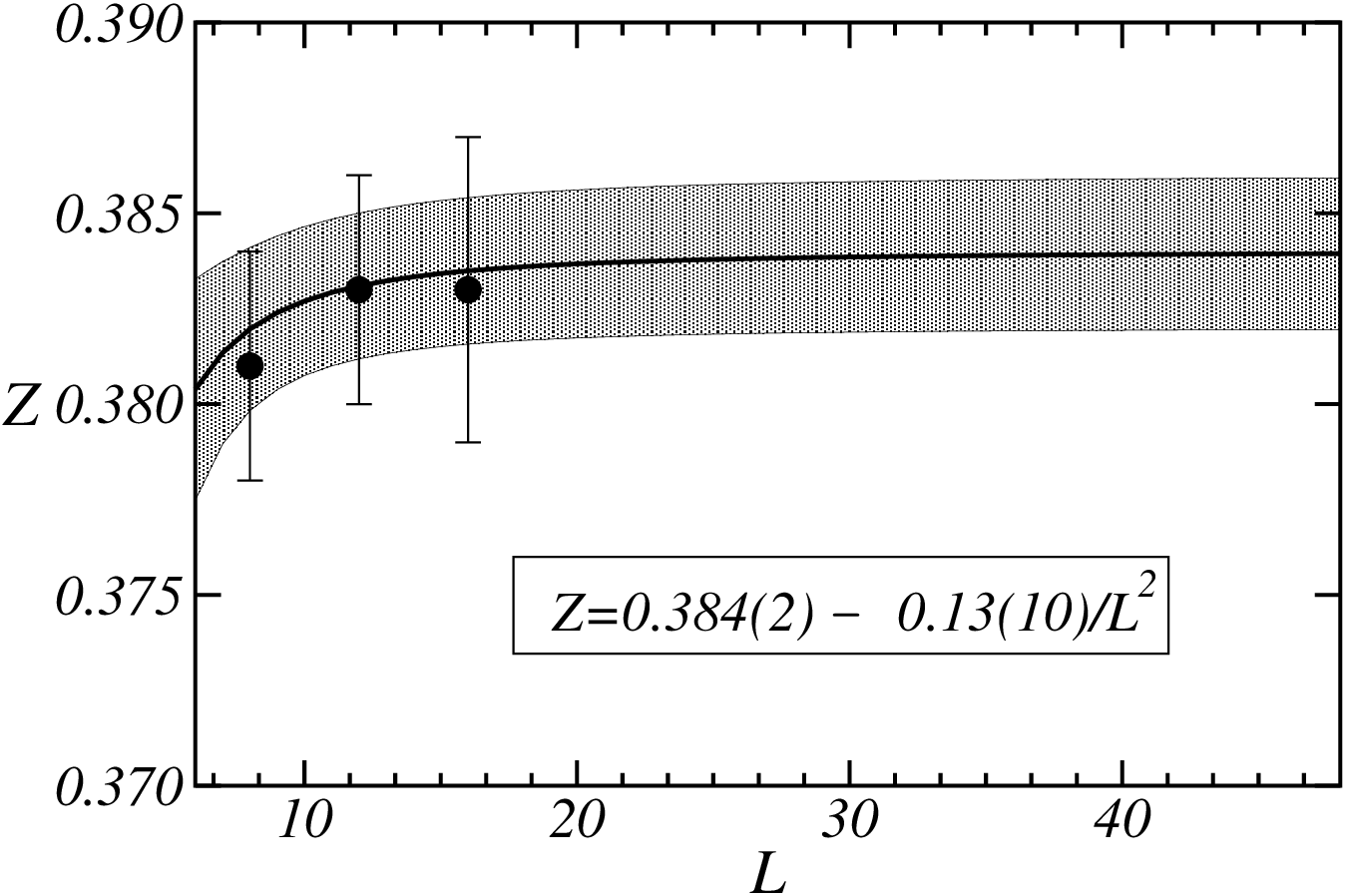}}
\vspace*{12pt}
\caption{$Z$ versus $L$ for the 1--smeared topological charge operator
$Q_L^{(1)}$ at $\beta=6.0$. The line is the result of the fit
(displayed in the legend) and the grey band is its 1--$\sigma$ error.}
\end{figure}

\vskip 1cm

As for the additive renormalization constant $M$ the procedure is
quite analogous. This time we calculate 
$M=\chi_L|_{Q=0}\equiv\langle\left(Q_L^{(1)}\right)^2\rangle/L^4|_{Q=0}$. 
Single trajectories consist again of 80 heat--bath steps with measurements
every 4 steps and cooling tests after each measurement. Thermalization
(with short distance fluctuations) require us to discard the initial
steps. In Table~2 the number of trajectories for each lattice
size is shown.

\newpage

In Figure~3 the results for $M$ extrapolated with a fit to the
form $A+ B/L^2$ are shown.
The nontrivial dependence on $L$ is evident.
Hitherto this dependence had not been detected because the statistics
(number of trajectories) was much lower than in the present paper.

\vskip 2cm

\begin{figure}
\centerline{\includegraphics[width=\columnwidth]{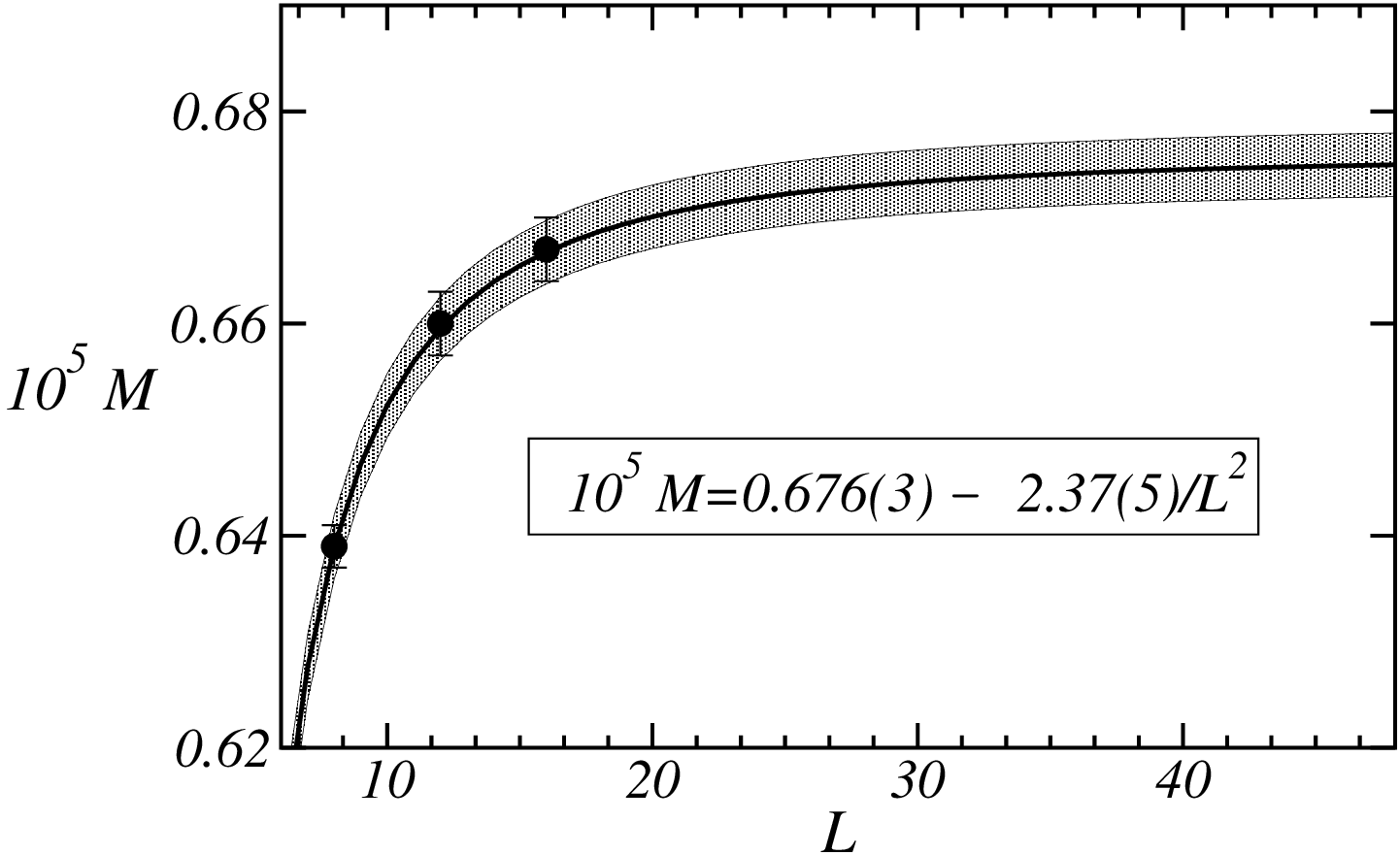}}
\vspace*{12pt}
\caption{$M$ versus $L$ for the 1--smeared topological charge operator
$Q_L^{(1)}$ at $\beta=6.0$. The line is the result of the fit
(displayed in the legend) and the grey band is its 1--$\sigma$ error.}
\end{figure}

\vskip 5mm

\noindent Again the $\chi^2/{\rm d.o.f.}$ test is rather small: 0.05. 
Also a nonzero cross correlation is present,
$\langle A B\rangle - \langle A\rangle \langle B\rangle
\approx -1.5\times 10^{-15}$
and the grey band of errors in Figure~3 was calculated by making use of it.

In Table~3 the results extrapolated for $Z$ and $M$ on the lattice sizes
used for the calculation of $\chi_P$ in section~2 are shown.

\newpage

\centerline{{\bf Table 3}: Extrapolated values for $Z$ and $M$.}
\vskip 1mm
{\centerline{
\vbox{\offinterlineskip
\halign{\strut
\vrule \hfil\quad $#$ \hfil \quad &
\vrule \hfil\quad $#$ \hfil \quad &
\vrule \hfil\quad $#$ \hfil \quad \vrule \cr
\noalign{\hrule}
{L^4} &
{Z} &
{10^5\; M} \cr
\noalign{\hrule}
16^4 &
0.383(2) &
0.667(3) \cr
\noalign{\hrule}
32^4 &
0.384(2) &
0.674(3) \cr
\noalign{\hrule}
48^4 &
0.384(2) &
0.675(3) \cr
\noalign{\hrule}
}}
}}

\vskip 1cm

It is interesting to check the results of $M$ by calculating it
on different background topological sectors. In fact this renormalization
constant can also be calculated as $M=\chi_L|_{Q=n} - Z^2 n^2/V$,
where $n$ can be any integer and $Z$ is the previously determined
renormalization constant. The results for $M$ obtained by
using different topological sectors $n$ agree within 
errors~\cite{vicaridg,delia}. Notice that this fact bears out the 
independence of $M$ on the topological charge sector.

In Figures~2 and~3 we have included 
the systematic error that is generated during the cooling test.
This error shows up because if some new instanton had appeared
during the updating process along the trajectory then this trajectory
must be discarded. However in some cases the cooling relaxation
eliminates this unwanted instanton and in this (rare(\footnote{This
event barely occurs because the autocorrelation time for the
topological charge is much larger than for other operators both in
the heating process~\cite{ddl} and in the cooling.}))
event the configuration is wrongly
taken as lying in the correct topological sector. This error tends
to modify the values of $M$ and $Z$.
When $\langle Q^2\rangle$ is calculated on the $n=0$ topological sector,
any jump from the $n=0$ sector enlarges the value of $M$.
On the other hand in the calculation of $Z$ 
on the $n=1$ sector, due to the symmetry of
the topological charge distribution around $n=0$, the configuration
prefers to move to the $n=0$ sector and 
this effect lowers the result for $Z$. Therefore this uncertainty
turns out to be asymmetric around the central value.

To estimate the influence of this error on the value of $M$ we 
measure also the plaquette. 

\begin{figure}
\centerline{\includegraphics[width=\columnwidth]{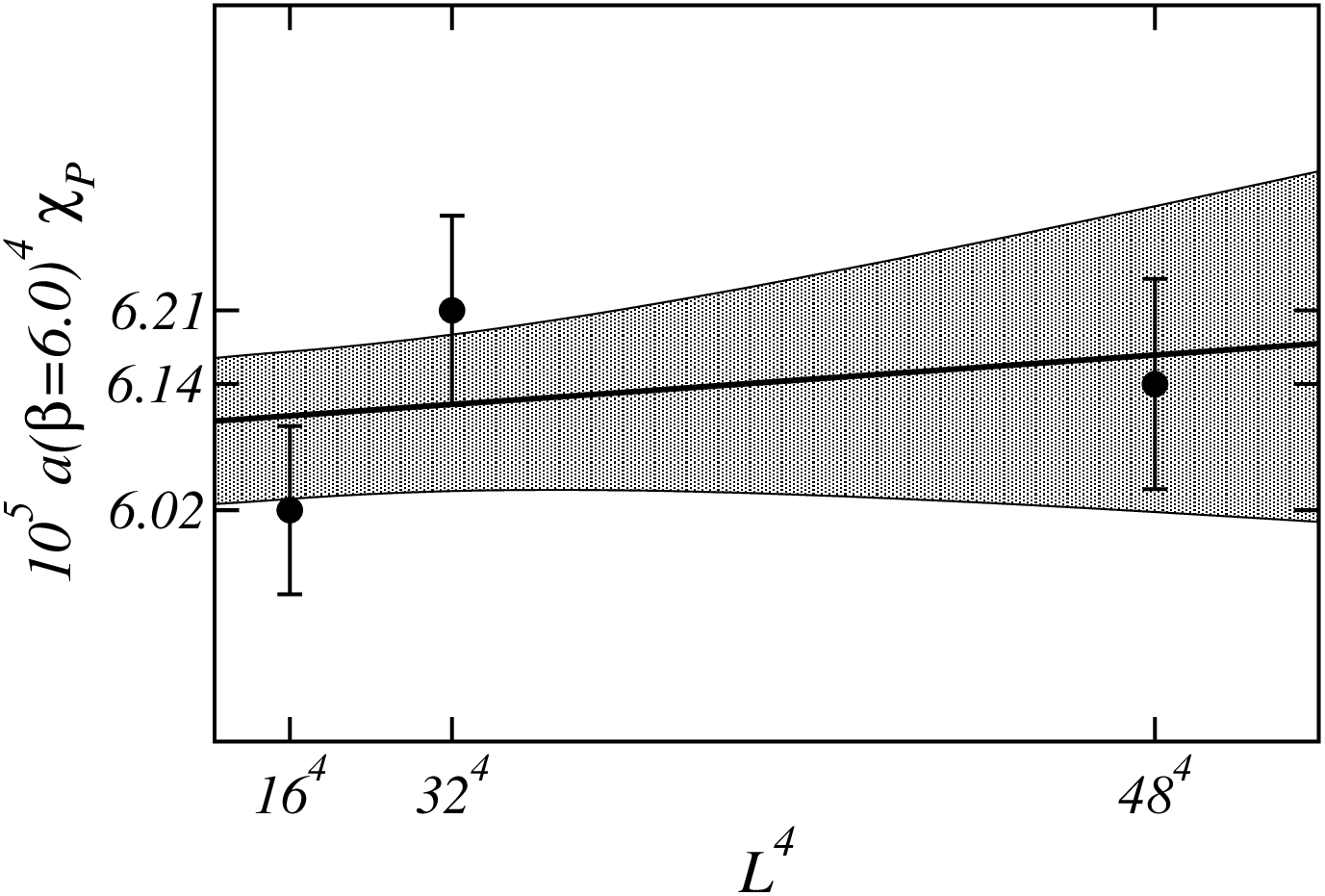}}
\vspace*{12pt}
\caption{$a^4\chi_P$ versus the lattice size for $\beta=6.0$.
The straight line is the result of the fit described in the text 
and the grey band is its 1--$\sigma$ statistical error.}
\end{figure}

\noindent After several heating steps the plaquette
thermalizes and a plateau in its signal appears.
The difference between
the value of $\chi_L|_{Q=0}$ at the step where this plateau sets in
and the value at the last step in the trajectory is taken as an estimate
of this error. An analogous  procedure was followed in the calculation of $Z$.
This error was added to the one corresponding to the choice of $\delta$
(see above) to form the total systematic error of our calculation.

We are planning to study this systematic error more carefully by
using the overlap algorithm~\cite{neuberger,luscher,giusti1,giusti2} 
in place of cooling to evaluate the 
background topological charge $Q$.

\section{Bound on $\alpha$ and calculation of $\chi$}

\vskip 5mm

With the data of Table~1 and~3 and Eq.(\ref{chilat}) we obtain
the values of $a^4\chi_P$ that are shown in Figure~4.
A fit (with the $\chi^2/{\rm d.o.f.}$ estimator) to the functional form 
$a^4\chi_P = \alpha^2 a^8 L^4 + a^4\chi$ yields
$\alpha^2 a^8 = 1.5(\pm 2.4)(^{+0.0}_{-2.0}) \; 10^{-13}$ and 
$a^4 \chi=6.09(\pm 0.07)(^{+0.16}_{-0.03})\;10^{-5}$; the first errors
being statistical and the second ones being systematic from cooling.

By using the values for $a(\beta)\Lambda_L$ and $T_c/\Lambda_L$ 
tabulated in~\cite{boyd} and the
ratio for $T_c/\sqrt{\sigma}$ from~\cite{biagio} we get 
$\Lambda_L=7.90(6)$ MeV and
$a(\beta=6.0)=0.1004(7)$~fermi. This leads to the 1--$\sigma$ bound on the
parity violating order parameter
\begin{equation}
\alpha \ltapprox  \left(\frac{1}{4\;{\rm fermi}}\right)^4 \;.
\label{albound}
\end{equation}

Much as an explicit $\theta$ term in the Lagrangian violates both
parity and time reversal and allows for a nonzero value of the
electric dipole moment $d_e$ of fermions, the spontaneous breaking of parity
driven by the topological charge operator brings about a
nonvanishing $d_e$. This $d_e$ is proportional to
$\alpha$ times some volume which in the case of the neutron 
can be estimated as the volume
occupied by this particle, roughly $1/m_N^4$. Moreover the whole 
effect would disappear
if the Lagrangian was chiral because then the extended theory
${\cal L}_{\rm QCD} + \theta Q(x)$ would be equivalent to ${\cal L}_{\rm QCD}$
with $\theta=0$. This implies that the derivative of the free energy
with respect to $\theta$ would vanish and spontaneous breaking of
parity could not occur. To lowest order, this means that the result
for $d_e$ must be proportional to the squared pion mass. 
Barring large numerical 
factors, using the neutron mass $m_N$ as the typical scale of the problem 
and assuming that the bound
in expression (\ref{albound}) applies to the 
full theory too, we estimate the
spontaneously generated electric dipole moment of the neutron as
\begin{equation}
d_e \approx e\, \frac{\alpha}{m_N^4} \frac{m_\pi^2}{m_N^3} 
    <  3.5\; 10^{-21} \; e\cdot{\rm cm}\;,
\label{dexp}
\end{equation}
to be compared with the experimental 
limit, $6.3\;10^{-26}\; e\cdot{\rm cm}$~\cite{harris}.
The result (\ref{dexp}) must not be confused with the electric dipole moment
calculated for the neutron in the presence of an explicit nonzero
physical value of $\theta$~\cite{baluni,veneziano1}. In our case 
the physical $\theta$ is zero.

We explicitely notice that our bound on the parameter $\alpha$
was obtained at fixed $\beta$ (fixed $a$) and it is not affected
by the continuum limit.

Our method also allows to extract the topological susceptibility
with great precision. From the value for $a^4\chi$ obtained in the
fit we get 
$\chi=\left(173.4(\pm 0.5)(\pm 1.2)(^{+1.1}_{-0.2})\;{\rm MeV}\right)^4$
where the first error is our statistical,
the second is the propagation of the error on $\Lambda_L$
and the third one is the systematic from cooling.
The two first errors must be added in quadrature.
This is the value of $\chi$ obtained at $\beta=6$. An
extrapolation $a\rightarrow 0$ should be made with comparable
precision. Usually the error derived from this extrapolation is negligible as
compared with the statistical one. However our value of the
topological susceptibility has been obtained after a very high statistics
simulation and this causes the systematic effects to become important.

The result for the topological susceptibility can be
compared with $\left(174(7)\;{\rm MeV}\right)^4$ obtained in~\cite{npb}
from a Monte Carlo simulation on a $16^4$ lattice.
It can also be compared with the analytical prediction~\cite{veneziano2}
$\chi\approx\left(180\;{\rm MeV}\right)^4$ obtained within a
$1/N_c$ expansion.

\vskip 2cm

\section{Conclusions}

\vskip 5mm

We have studied the consequences of the spontaneous breaking of parity and
time inversion symmetries on the topological charge operator 
in pure Yang--Mills theory with gauge group $SU(3)$. We assumed that
in the infinite volume limit the free energy $E(\theta)$ of the extended
theory  ${\cal L}_{\rm QCD} + \theta {\cal O}$ has a cusp at
its minimum, $\theta=0$ that signals a nonzero value of the
topological charge in the vacuum, d$\langle Q\rangle/$d$V =\pm\alpha$.

By calculating the value of the topological
susceptibility on the lattice to high precision 
we have obtained a bound on $\alpha$: within
1--$\sigma$ error, there is no spontaneously generated net topological
charge in a volume of (4 fermi)$^4$. The bound on $\alpha$ becomes
an upper bound for the neutron electric dipole moment which however
is 4 to 5 orders of magnitude less precise than the corresponding 
experimental limit.

As a byproduct of our Monte Carlo simulation, we have also obtained
a precise result for the topological susceptibility,
$\chi=\left(173.4(\pm 0.5)(\pm 1.2)(^{+1.1}_{-0.2})\;{\rm MeV}\right)^4$
(the errors are respectively statistical
from our Monte Carlo simulation, statistical from the 
value used for $\Lambda_L$ and systematic from the cooling used during the
evaluation of the renormalization constants).
The precision in the statistics forced us to study this systematic
error which otherwise can be neglected. 
The continuum limit should be made with comparable precision.

We have done the calculation by using the so--called ``field
theoretical method'' where the renormalization constants relating
the lattice and the physical susceptibilities are explicitely computed.
This method proves to be fast and efficient enough to allow to obtain
a huge number of measurements on rather large lattices, (see
Table~1). The APEmille facility in Pisa was used for the runs.

The renormalization constants 
in~Eq.(\ref{chilat}), $Z$ and $M$, have been calculated 
with high statistics at various
volumes. The value of $Z$ looks rather stable with the volume,
while the value of $M$ displays a clean volume dependence (see Fig.~3).

\end{document}